\documentclass[conference, 9pt]{IEEEtran}
\IEEEoverridecommandlockouts
\usepackage{multirow}

\usepackage[utf8]{inputenc} % allow utf-8 input
\usepackage[T1]{fontenc}    % use 8-bit T1 fonts
\usepackage{url}            % simple URL typesetting
\usepackage{booktabs}       % professional-quality tables
\usepackage{amsfonts}       % blackboard math symbols
\usepackage{nicefrac}       % compact symbols for 1/2, etc.
\usepackage{microtype}      % microtypography
\usepackage{xcolor}         % colors

\usepackage{amsmath}
\usepackage{mathtools}
\usepackage{graphicx}
\usepackage{caption}
\usepackage{subcaption}
\usepackage{pgfplots}
\usepackage{url}
\usepackage{tikz}
\usepackage{hyperref}
\newcommand{\Reals}{\mathbb R}

\newcommand{\Output}{\mathbf H}
\newcommand{\OutputVector}{\mathbf h}
\newcommand{\Input}{\mathbf X}
\newcommand{\InputVector}{\mathbf x}

\newcommand{\SummaryMean}{\bar{\mathbf s}}

\newcommand{\Transformation}{f}

\newcommand{\Summary}{s}

\newcommand{\CombineFunction}{c}

\newcommand{\Time}{t}
\newcommand{\OtherTime}{u}
\newcommand{\ChunkSize}{C}
\newcommand{\LeftContextSize}{L}
\newcommand{\NumVisibleFrames}{\bar{T}_t}
\newcommand{\Mask}{\mathbf M}
\newcommand{\MaskElement}{m}
\newcommand{\Length}{T}

\newcommand{\InputDimensionality}{D}
\newcommand{\OutputDimensionality}{D'}
\newcommand{\TransformedDimensionality}{D''}
\newcommand{\SummarySize}{D'''}

\newcommand{\Complexity}{\mathcal{O}}

\begin{document}

% title here must exactly match the title entered into the paper submission system

\def\BibTeX{{\rm B\kern-.05em{\sc i\kern-.025em b}\kern-.08em
    T\kern-.1667em\lower.7ex\hbox{E}\kern-.125emX}}

\title{Linear Time Complexity Conformers with SummaryMixing for Streaming  Speech Recognition}

\author{\IEEEauthorblockN{Titouan Parcollet, Rogier van Dalen, Shucong Zhang, Sourav Batthacharya}
\IEEEauthorblockA{\textit{Samsung AI Center - Cambridge} \\
Cambridge, United Kingdom \\
\{t.parcollet,r.vandalen,s1.zhang,sourav.b1\}@samsung.com}
}

\maketitle

\begin{abstract}
Automatic speech recognition (ASR) with an  encoder equipped with self-attention, whether streaming or non-streaming, takes quadratic time in the length of the speech utterance. This slows down training and decoding, increase their cost, and limit the deployment of the ASR in constrained devices. SummaryMixing is a promising linear-time complexity alternative to self-attention for non-streaming speech recognition that, for the first time, preserves or outperforms the accuracy of self-attention models. Unfortunately, the original definition of SummaryMixing is not suited to streaming speech recognition. Hence, this work extends SummaryMixing to a Conformer Transducer that works in both a streaming and an offline mode. It shows that this new linear-time complexity speech encoder outperforms self-attention in both scenarios while requiring less compute and memory during training and decoding.
\end{abstract}
\section{Introduction}

Streaming and non-streaming automatic speech recognition (ASR) have followed the general trend in deep learning and increased steadily both in model size and architecture complexity. A few speech encoders have reached billions of neural parameters \cite{radford2022robust} while improving the recognition accuracies. As a consequence of this success, ASR systems are now part of numerous real-life products \cite{nassif2019speech}. Such an increase in required computing and memory resources, however, is in conflict with the surging demand for constrained-resources deployment \cite{arasteh2016iot} of on-device ASR models. Streaming ASR is a great example as the perceived latency at decoding time from the end-user is almost as important as the word error rate (WER). 

Multi-head self-attention (MHSA) \cite{vaswani2017attention} is one of the core components of most ASR systems. Its success lies in the ability to capture fine-grained global interactions between pairs of acoustic frames. In an offline, or non-streaming ASR scenario, self-attention will be computed over the whole speech utterance \cite{kim2023branchformer}. For streaming, the self-attention output of a given time step often depends only on all or a subset of the past frames. Future context can also be added via chunked streaming ASR as a few future frames can be cached and attended by self-attention, at the cost of a slightly higher latency \cite{li2021better,li2023dynamic}. Unfortunately, MHSA complexity grows quadratically with the input sequence length. This limits in practice, for memory and run-time reasons, the number of past frames that may be used by self-attention, reducing the recognition accuracy compared to an infinite past context. Hence, it appears critical to find an alternative to self-attention that may deal without issue with a growing and infinite past contact for both accuracy and efficiency reasons.

Typical efficient changes to self-attention outside of the speech modality are low-rank approximation \cite{tay2022efficient}, linearization \cite{wang2020self}, or sparsification \cite{child2019generating}. For ASR more specifically, the Squeezeformer \cite{kim2022squeezeformer}, the Efficient Conformer \cite{burchi2021efficient} and the Emformer \cite{shi2021emformer} aim to reduce the impact of the quadratic complexity by reducing the sequence length. Using these methods, the memory and training time are indeed reduced, but the quadratic time complexity remains. Instead, we aim to remove entirely the quadratic time complexity. In that direction, the Fastformer \cite{wu2021fastformer} proposed to linearise self-attention. It was successfully applied to non-streaming ASR exhibiting faster training times but at the cost of slightly degraded accuracies against MHSA \cite{peng2022branchformer}. 

More recently, a few articles reported that under certain conditions, pair-wise MHSA operations in trained ASR models including transformers and conformers, behave like simple feed-forward layers \cite{zhang2021usefulness, shim2022understanding}. This observation led to the development of SummaryMixing for ASR \cite{parcollet2024summarymixing}, an alternative to MHSA that takes linear time in the sequence length. For the first time, and in non-streaming ASR scenarios, SummaryMixing-equipped state-of-the-art models including conformers \cite{gulati2020conformer} and branchformers \cite{peng2022branchformer} reduced significantly the memory and training time while outperforming both MHSA and Fastformer on achieved WERs. These findings have also been validated with self-supervised learning (SSL) and different downstream speech tasks including ASR, speaker verification, intent classification and emotion recognition \cite{zhang2024linear}. In short, SummaryMixing summarises the whole sequence in a single vector modelling global relations between frames. This summary vector is then concatenated back to each frame of the input sequence. Frames are also individually transformed non-linearly to model local interactions. Unfortunately, the definition of SummaryMixing is not compatible with streaming ASR. 

This article extends SummaryMixing to streaming ASR (Section \ref{sec:method}) and evaluates its performance in a unified streaming and non-streaming training and decoding framework using dynamic chunk training (DCT) and dynamic chunk convolution (DCCONV) \cite{li2023dynamic} on two ASR datasets (Section \ref{sec:exps}). The behavior of SummaryMixing is then compared to MHSA with varying sequence lengths and an infinite past context in terms of inference real-time factor, WERs, and memory consumption. Experiments conducted with a state-of-the-art streaming and non-streaming conformer Transducer show that SummaryMixing achieves equivalent or better WER than MHSA while reducing the training and decoding time as well as the memory footprint both during training and inference. The code is released as a recipe of the SpeechBrain toolkit%
\footnote{Code is available under the licence CC-BY-NC 4.0 at: \url{https://github.com/SamsungLabs/SummaryMixing}}. We also release in the SpeechBrain toolkit a half-precision Transducer loss effectively halving the peak memory required during training.

\section{Streaming and Non-Streaming SummaryMixing}
\label{sec:method}

This section introduces all the necessary components to perform streaming and non-streaming ASR (section \ref{subsec:dct}) with a conformer transducer architecture and SummaryMixing (Section \ref{subsec:summarymixing}).

\subsection{Dynamic Chunk Convolutions and Training}
\label{subsec:dct}

Dynamic chunk training (DCT) \cite{yao21_interspeech} makes an ASR system capable of operating both in a streaming and offline settings. This is achieved by exposing the ASR model to a variety of context lengths, or \textit{dynamic chunks}, during training. Indeed, in the streaming setting, the model cannot be allowed to look ahead far as small chunks of audio are processed at a time. Conversely, for offline ASR, the model can process the whole speech utterance. 

What the model is allowed to see can be expressed as a mask $\Mask \in \{0,1\}^{\Length \times \Length}$ with $\Length$ the number of frames or time steps in the sequence. The element $\MaskElement_{\Time, \OtherTime}$ of the mask indicates whether at time step $\Time$ the model is allowed to see the input, from time $\OtherTime$.
The whole batch uses the same mask, which is constructed as follows.
A chunk size $\ChunkSize$ is drawn randomly from $[1, \Length]$.
To restrict the quadratic complexity of self-attention, the left context is often limited to $\LeftContextSize$ chunks, with $\LeftContextSize$ drawn from $\big[0, \lceil\frac{\Length}{\ChunkSize}\rceil\big]$.
The mask is then set to:
\begin{align}
    \MaskElement_{\Time, \OtherTime} &=
        \left\{\begin{array}{ll}
            1 & \text{if~}
                \left\lfloor \frac{\OtherTime}{\ChunkSize} \right\rfloor - \LeftContextSize
                \leq
                \left\lfloor \frac{\Time}{\ChunkSize} \right\rfloor
                \leq
                \left\lfloor \frac{\OtherTime}{\ChunkSize} \right\rfloor
            ;\\
            0 & \text{otherwise.}
        \end{array}\right.
\end{align}
When $\ChunkSize = \Length$, the model can see the complete input, since then $\MaskElement_{\Time, \OtherTime} = 1$ for all $\Time$ and $\OtherTime$.
Otherwise, setting $
    \LeftContextSize = \lfloor \Length / \ChunkSize \rfloor
$ results in infinite left context.
If the left context is infinite, then the mask only grows as time proceeds: $\MaskElement_{\Time, \OtherTime} \leq \MaskElement_{\Time+1, \OtherTime}$. The implementation implications of this chunking depend on the type of neural network.

Convolutional layers require a specific design to avoid data leakage from future frames. If using standard convolutional layers, kernels will be trained to see a few frames after the chunk boundaries, and therefore performance will degrade significantly at inference time as these frames are not available. Causal convolutions \cite{yao21_interspeech} address this issue by shifting to the left of the convolutional kernels with the current frame being the rightmost element. However, in the case of DCT, it also means that every frame computed with causal convolution does not integrate information from the future within the chunk boundaries, resulting in performance degradation \cite{li2023dynamic}. Dynamic chunk convolution \cite{li2023dynamic} (DCCONV) addresses all these issues by keeping standard convolutional kernels centered on the current frame, hence allowing within-chunk future information, but with masking according to the boundaries of the chunk. Hence, convolutional kernels will see different levels of masking as they progress on the rightmost frames.  

For self-attention layers, things are much simpler as the attention score is masked according to the chunk boundaries. Hence, no frames from outside of the chunk or the left context are being attended by the attention mechanism.

ASR systems trained with DCT and DCCONV are capable of both streaming and non-streaming ASR without architectural change or fine-tuning as shown in \cite{li2023dynamic}. We rely on the DCT available in SpeechBrain \cite{speechbrain}.

\subsection{Streaming SummaryMixing}
\label{subsec:summarymixing}

\begin{figure}

    \center
    \begin{subfigure}[b]{0.43\columnwidth}
        \includegraphics{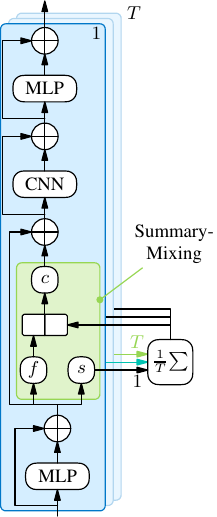}
    \end{subfigure}%
    \hspace{4mm}
    \begin{subfigure}[b]{0.45\columnwidth}
        % Note: The right figure is drawn slightly on top of the other, so that
        % the "SummaryMixing" label can be re-used.
        \hspace*{-5mm}
        \includegraphics{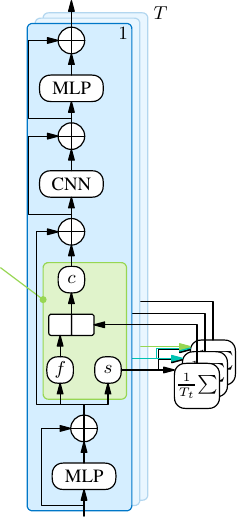}
    \end{subfigure}
    \caption{The Conformer with non-streaming (left) and streaming (right) SummaryMixing.}
    \label{fig:conformer}
\end{figure}

SummaryMixing was introduced in \cite{parcollet2024summarymixing} as a drop-in replacement for self-attention in offline ASR that reduces the time complexity from quadratic to linear. In the following, we recall its basics, extend it to the streaming ASR scenario, and introduce it to the conformer transducer.

Much like self-attention, SummaryMixing takes an input sequence $\Input \in \Reals^{\Length \times \InputDimensionality} = \{\InputVector_0,\ldots,\InputVector_{\Length}\}$ of $\Length$ feature vectors $\InputVector_{\Time}$ of length $\InputDimensionality$, and transforms them into hidden representations $\Output \in \Reals^{\Length \times \OutputDimensionality} = \{\OutputVector_0,\ldots,\OutputVector_{\Length}\}$.

In particular, SummaryMixing summarises a whole utterance in a single vector with a linear-time complexity summary function. This single vector is then passed back to every time step of the sequence. First, the input $\InputVector_{\Time}$ is transformed by two non-linear functions e.g. two dense layers with activation functions. One is the local transformation function $\Transformation: \Reals^{\InputDimensionality} \rightarrow \Reals^{\TransformedDimensionality}$ while the second is the summary function $\Summary: \Reals^{\InputDimensionality} \rightarrow \Reals^{\SummarySize}$. The output of the latter function is then averaged over time ($\frac1{\Length}\!\sum$) to obtain the mean vector $\SummaryMean$. This single vector is concatenated back to each time step of the output of the local transformation ($\Transformation$). This concatenation is processed by the non-linear combiner function $\CombineFunction: \Reals^{\TransformedDimensionality + \SummarySize} \rightarrow \Reals^{\OutputDimensionality}$. Computing the mean from $\SummaryMean$ takes $\Complexity(\Length)$ time, compared to the usual quadratic cost of MHSA. SummaryMixing can be formally summarised as:

\begin{align}
    \SummaryMean &= \frac1{\Length} \sum_{\Time=1}^{\Length} \Summary(\InputVector_{\Time})
    ;&
    \OutputVector_{\Time} &= \CombineFunction(\Transformation(\InputVector_{\Time}), \SummaryMean).
    \label{eq:summarymixing}
\end{align}%

Extending SummaryMixing from \eqref{eq:summarymixing} to streaming ASR means that the whole sentence will not be summarised into a single vector anymore.
Instead, as Figure \ref{fig:conformer} illustrates, the average should be computed over only the frames that are visible to the model.
The summary vector, written $\SummaryMean_t$, is specific to the time step.
The number of visible frames at time $\Time$ can be computed  as $
    \NumVisibleFrames = \sum_{\OtherTime=1}^{\Length} \MaskElement_{\Time, \OtherTime}
$.
Streaming SummaryMixing can then be described by changing \eqref{eq:summarymixing} to:
\begin{align}
    \SummaryMean_t &= \frac1{\NumVisibleFrames} \sum_{\OtherTime=1}^{\Length} \MaskElement_{\Time, \OtherTime} \cdot \Summary( \InputVector_{\OtherTime})
    ;&
    \OutputVector_{\Time} &= \CombineFunction(\Transformation(\InputVector_{\Time}), \SummaryMean_t)
    .
    \label{eq:streaming_summarymixing}
\end{align}
When the left context is infinite, the mask only grows, and the sum $\sum_{\OtherTime=1}^{\Length} \cdot$ can be computed incrementally, keeping time complexity linear compared to quadratic for MHSA. SummaryMixing therefore makes the use of an infinite left context, i.e. better recognition accuracies, more practical than MHSA. For DCT, the sum takes on a new value only when a new chunk comes in. Conversely to MHSA, SummaryMixing can easily use an infinite left context.\\ 

\begin{table*}[ht!]
    \centering
    \caption{Speech recognition word error rates (WER) with a non-streaming and streaming conformer transducer with chunks of 1280ms, 640ms and 320ms trained with and without dynamic chunk training. The left context is infinite. }
    \label{tab:res_asr}
    \begin{tabular}{l@{}c|@{~}c@{~}c@{~}c@{~}|@{~}c@{~}c@{~}c@{~}|@{~}c@{~}c@{~}c@{~}|@{~}c@{~}c@{~}c@{~}|c@{~~}c@{~~}|c@{~~}c@{~~}}
    \toprule
        && \multicolumn{3}{c|@{~}}{\textbf{Non-streaming}} & \multicolumn{3}{c|@{~}}{\textbf{1280ms}} & \multicolumn{3}{c|@{~}}{\textbf{640ms}}  & \multicolumn{3}{c|}{\textbf{320ms}}  & \multicolumn{4}{c}{\textbf{Training}} \\
        & \textbf{Dynamic} & \multicolumn{3}{c|@{~}}{\textit{WER \%}} & \multicolumn{3}{c|@{~}}{\textit{WER \%}} & \multicolumn{3}{c|@{~}}{\textit{WER \%}}  & \multicolumn{3}{c|}{\textit{WER \%}}  & \multicolumn{2}{c|}{\textbf{GPU}} & \multicolumn{2}{c}{\textbf{VRAM}} \\
        & \textbf{chunk} & \textit{dev-} & \textit{test-} & \textit{test-} & \textit{dev-} & \textit{test-} & \textit{test-} & \textit{dev-} & \textit{test-} & \textit{test-}  & \textit{dev-} & \textit{test-} & \textit{test-}  & \multicolumn{2}{c|}{\textit{hours}} & \multicolumn{2}{c}{\textit{GB}} \\
        \textbf{Librispeech} & \textbf{training}
        & \textit{clean} & \textit{clean} & \textit{other} & \textit{clean} & \textit{clean} & \textit{other} & \textit{clean} & \textit{clean} & \textit{other}  & \textit{clean} & \textit{clean} & \textit{other}  & \textit{fp16} & \textit{fp32} & \textit{fp16} & \textit{fp32} \\
    \midrule
           Self-attention & --- & 2.7 & 3.0 & 7.0 & --- & --- & --- &  --- & --- & --- & --- & --- & --- & 29.0 & 39.5  & 20.1  & 40.2   \\
           SummaryMixing & ---  & 2.7 & 2.9 & 7.0 & --- & --- & --- &  --- & --- & --- & --- & --- & --- & 24.5 & 34.7  & 16.9  & 32.9   \\
           Self-attention & \checkmark & 3.0 & 2.9 & 7.0 & 3.1 & 3.2 & 8.0  & 3.2 & 3.4 & 8.6 & 3.4 & 3.7 & 9.6 & 29.7 & 40.1  & 20.2 & 40.2   \\
           SummaryMixing & \checkmark & 2.9 & 2.9 & 7.0 & 2.9 & 3.2 & 8.1  & 3.1 & 3.4 & 8.6 & 3.3 & 3.7 & 9.6 & 25.0 & 35.1  & 16.9 & 33.0  \\
    \midrule
    \midrule
    \textbf{Voxpopuli} && \multicolumn{3}{c|@{~}}{\textit{Dev.\quad Test}} & \multicolumn{3}{c|@{~}}{\textit{Dev.\quad Test}} & \multicolumn{3}{c|@{~}}{\textit{Dev.\quad Test}} & \multicolumn{3}{c|}{\textit{Dev.\quad Test}}  &  &  \\
    \midrule
           Self-attention & ---  & \multicolumn{3}{c|@{~}}{11.5 \quad 11.7} & \multicolumn{3}{c|@{~}}{--- \quad~~ ---} & \multicolumn{3}{c|@{~}}{--- \quad~~ ---} & \multicolumn{3}{c|}{--- \quad~~ ---} & 8.0 & 12.9  & 36.0 & 71.1 \\
           SummaryMixing & ---  & \multicolumn{3}{c|@{~}}{11.0 \quad 11.2} & \multicolumn{3}{c|@{~}}{--- \quad~~ ---} & \multicolumn{3}{c|@{~}}{--- \quad~~ ---} & \multicolumn{3}{c|}{--- \quad~~ ---} & 7.1 & 10.0  & 29.1 & 61.3 \\
           Self-attention & \checkmark & \multicolumn{3}{c|@{~}}{11.6 \quad 11.7} & \multicolumn{3}{c|@{~}}{12.3 \quad 12.6} & \multicolumn{3}{c|@{~}}{13.0 \quad 13.4} & \multicolumn{3}{c|}{14.6 \quad 14.6} & 8.4 & 13.3  & 36.1 & 71.1  \\
           SummaryMixing & \checkmark & \multicolumn{3}{c|@{~}}{11.2 \quad 11.3} & \multicolumn{3}{c|@{~}}{12.1 \quad 12.4} & \multicolumn{3}{c|@{~}}{12.8 \quad 13.0} & \multicolumn{3}{c|}{14.1 \quad 14.2} & 7.3 & 10.4  & 29.2 & 61.3  \\
    \bottomrule
    \end{tabular}
\end{table*}

\noindent\textbf{Conformer Transducers with SummaryMixing.} The conformer-based transducer is a popular architecture for streaming ASR \cite{burchi2021efficient} mainly due to its capability to generate partial hypotheses without processing the whole input sequence. Making the conformer transducer compatible with SummaryMixing is straightforward, as only the acoustic encoder i.e. the conformer part, needs to be updated. Following \cite{parcollet2024summarymixing}, and based on the fact that self-attention and SummaryMixing are expected to fulfill the same role within the encoder, we replace every MHSA cell in the conformer with a SummaryMixing one, hence leading to an ASR architecture without any attention. Convolutional layers are also changed to dynamic chunck convolutions to allow DCT \cite{li2023dynamic}.  The integration of SummaryMixing to the conformer encoder is depicted in Figure \ref{fig:conformer}. \\

% \noindent\textbf{Accelerating the input projections.} In the original SummaryMixing \cite{parcollet2024summarymixing}, both $\Summary$ and $\Transformation$ are implemented as multilayer perceptions or deep neural networks. We found experimentally that a single linear transformation was sufficient to reach the optimal accuracy threshold. Therefore, and to accelerate even further the training, $\Summary$ and $\Transformation$ can be merged in a single linear layer with an output size combining the output dimension of both functions, hence starting a single larger CUDA kernel rather than two smaller ones.  

\section{Experiments}
\label{sec:exps}

The effectiveness of SummaryMixing for streaming and non-streaming ASR is first evaluated on two datasets (section \ref{subsec:asr_res}). Then, we provide an analysis of the efficiency gains in terms of real-time factor and peak memory consumption as well as on the impact of the audio duration on the word error rate (section \ref{subsec:extended_analysis}).

\subsection{Experimental protocol}
\label{subsec:exp_protocol}

SummaryMixing has already been shown to outperform many baselines for non-streaming ASR \cite{parcollet2024summarymixing}. Hence, we decided to compare it to the fastest available implementation of a conformer-transducer for streaming and offline ASR with MHSA on SpeechBrain\footnote{SpeechBrain version 1.0.0 is used for all experiments.}.\\ 

\noindent\textbf{Speech recognition datasets.} Librispeech \cite{panayotov2015librispeech} and the English set of Voxpopuli \cite{wang-etal-2021-voxpopuli} are considered as benchmark datasets. On Librispeech, the tokenizer and the recurrent neural language model (RNNLM) are pre-trained and originate from the official SpeechBrain recipe. The full 960 hours are used for training while results are reported on the standard validation and test sets. Voxpopuli is a multilingual dataset containing recordings from the European Parliament. The acoustic conditions including recording hardware, background noises, and speech accents are more challenging than Librispeech. In practice, we removed all sentences over 100 seconds to fit the memory budget with MHSA while training. The training set contains 522 hours of speech while the validation and test sets are made of 5 hours. No language model is used for Voxpopuli. \\

\noindent\textbf{Architecture and training details.} Hyperparameters are almost identical to the DCT recipe from SpeechBrain$^2$ and are open-sourced in our repository$^1$. In brief, the Transducer architecture is made of a 12-block conformer acoustic encoder equipped either with MHSA or SummaryMixing and DCCONV, a one-layered LSTM predictor network, and a joiner with a single hidden layer. The joiner performs a sum before the projection. The total number of trainable parameters for the models is 80M. Multi-task learning is performed for the first 10 epochs with the CTC loss \cite{graves2012connectionist}. Models are trained for 90,000 and 50,000 steps on Librispeech and Voxpopuli respectively with a warmup (25,000 steps) and exponential decay learning rate scheduler. The learning rate peaks at 0.0008. Dynamic chunk training parameters are strictly identical to the original recipe with a streaming probability of 0.6 and chunks ranging from 320ms to 1280ms. The left context also randomly varies from 320ms to 1280ms. Full context training is performed 40\% of the time enabling the model to behave properly when an infinite left context is given. The output vocabulary size is 1,000 for both datasets. Batches are created via dynamic batching and a total batch duration of 900 seconds. Models were trained on three A40 46GB. Each GPU was able to process 150 seconds of speech, hence a gradient accumulation of two was applied to reach the 900 seconds per batch.\\

\noindent\textbf{Half-precision and transducers.} Most existing toolkits including SpeechBrain \cite{speechbrain}, ESPnet \cite{watanabe2018espnet} or k2 \cite{yao2023zipformer} integrate half-precision training for the neural network parts of transducer systems greatly boosting the training speed. In ESPnet or SpeechBrain, however, the memory footprint remains unchanged due to a large bottleneck when instantiating the four-dimensional tensor merging time-steps and tokens of the transducer. Indeed, the loss computation is computed in single precision and the large tensor must be cast back to this type, creating a spike in memory requirement. To avoid this casting operation, and to divide by a factor of two the memory consumption, we contributed to SpeechBrain a fully half-precision transducer loss computation using Numba \cite{lam2015numba}. 

\subsection{Speech recognition results}
\label{subsec:asr_res}

The effectiveness of SummaryMixing for streaming ASR is first evaluated on Librispeech and Voxpopuli. Table \ref{tab:res_asr} reports the WER for SummaryMixing and MHSA models once trained with and without DCT and evaluated on a non-streaming scenario or with streaming chunks of length 1280 ms, 640 ms, and 320 ms. In the streaming setting, the left-context is infinite. It also reports the peak VRAM for the full training as well as the total number of GPU hours needed. 

\begin{figure*}[!t]
\centering

\includegraphics{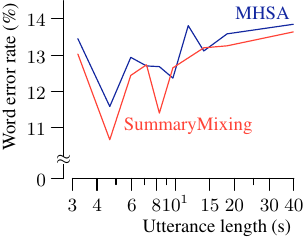}%
\hspace*{\stretch{1}}%
\includegraphics{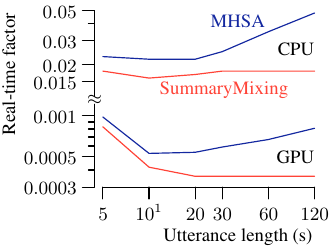}%
\hspace*{\stretch{1}}%
\includegraphics{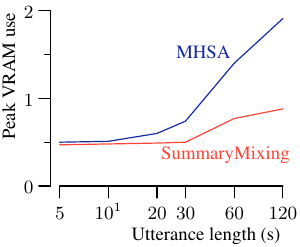}

\caption{Word error rate variations of the MHSA- and SummaryMixing- equipped models trained on VoxPopuli depending on the utterance length (left curve). Sentences are obtained from the test set of VoxPopuli. The middle curve shows the inference real-time factor observed with CPU and GPU for the two models. The right-most curve gives the peak VRAM from both models when decoding. The left context is infinite for all experiments.} \label{fig:curves}
\end{figure*}

SummaryMixing either outperforms or matches the WER of MHSA with both datasets and on all scenarios. If decoding is performed without streaming, SummaryMixing WER is lower than MHSA by 0.1\% absolute on the ``dev-clean'' set and identical on the two others. When streaming with chunks of length 640 ms and 320 ms the same 0.1\% absolute WER improvement over MHSA is observed on ``dev-clean'' for SummaryMixing. At 1280 ms, however, SummaryMixing improves by 0.2\% absolute on ``dev-clean'' but degrades by 0.1\% absolute on ``test-clean''. This is the only degradation observed with SummaryMixing on both datasets and across all scenarios. In particular, with VoxPopuli, SummaryMixing always outperforms MHSA. The average absolute WER reduction obtained by SummaryMixing is 0.34\%. This difference increases with the reduction of latency as the streaming scenario with a 320 ms chunk size sees SummaryMixing performing at a WER of 14.1\% against 14.6\% for MHSA on the validation subset. 

On average, SummaryMixing-equipped conformer transducers are 15.5\% faster to train than MHSA ones. Memory-wise, and despite the rather short sentences contained in Librispeech, the peak VRAM using mixed precision goes from 20.2 GB for MHSA to 16.9 GB, hence a 16\% decrease in memory budget. On VoxPopuli, with longer utterances, this increases to 19\%. SummaryMixing therefore offers better or similar performance for non-streaming and streaming ASR, but at a significantly lower training cost. Finally, our half-precision Transducer loss divides by more than two the required VRAM. 

\subsection{Extended analysis of streaming SummaryMixing}
\label{subsec:extended_analysis}

An extended analysis of the characteristics of SummaryMixing is conducted along three axes: encoder inference real-time factor (RTF) analysis on CPU and GPU, peak memory usage, and WER variations given the sequence length. Indeed, and due to its averaging mechanism, it sounds plausible that SummaryMixing is more impacted than MHSA when the available context increases. It is critical to note that the left-context is infinite for all measurements. Indeed, an infinite left context leads to higher recognition accuracies. In that scenario, SummaryMixing has a linear-time complexity while self-attention time complexity is quadratic. MHSA could also exhibit a linear-time complexity by reducing the left context to a fixed number of chunks, but the WER would then be negatively impacted. However, our goal is to maximize the accuracy and minimize the latency and compute requirements. Figure \ref{fig:curves} shows this analysis.\\

\noindent\textbf{WER given utterance length.} The 5 hours of speech contained in the test set of VoxPopuli are grouped into 10 buckets of increasing length. We then compute the WER of all the sentences in each bucket and report their value on the left-most curves of Figure \ref{fig:curves} for MHSA and SummaryMixing. Streaming decoding from models trained with VoxPopuli is performed with a chunk size of 640 ms, hence corresponding to the middle column of Table \ref{tab:res_asr}. While the WER increases slightly with the utterance length for both MHSA and SummaryMixing, the rate of increase is not higher for SummaryMixing. Therefore, the averaging mechanism of SummaryMixing does not seem to impact adversaly the WER with longer sequences compared to MHSA in this setting.\\

\noindent\textbf{Encoder RTF analysis.} The RTF is obtained at the acoustic encoder level, i.e. the conformer, as all the other blocks are the same both for MHSA and SummaryMixing. The decoding algorithm also is excluded as it does not affect the comparison and is usually an order of magnitude slower than the inference. Instead, we wish to focus on the impact of SummaryMixing over MHSA on the encoder when inferring. We generate five sets of 100 random input sequences sampled at 16 kHz of length $[5,10,20,30,60,120]$ seconds and pass them throughout the acoustic encoder. The random nature of the input does not affect the measured RTF as no autoregressive behavior is included in the measurements. RTF are measurements either on an isolated Nvidia A40 46GB GPU or on four cores of an Intel Xeon 5218 CPU. The middle curve of Figure \ref{fig:curves} shows the outcomes of that experiment. It clearly appears that the RTF, both on CPU and GPU, of the MHSA-equipped encoder degrades with the increase of the utterance length. This is not true for SummaryMixing as the RTF remains constant due to its linear-time complexity. SummaryMixing always is faster to infer compared to MHSA.\\

\noindent\textbf{Memory consumption.} We also store the peak VRAM usage during the GPU RTF analysis experiments to generate curves of memory requirements over the ten different sets of utterance lengths. The right-most curve of Figure \ref{fig:curves} reports the results. MHSA suffers from a much faster increase in memory consumption than SummaryMixing given the sentence length as peak VRAMs of 0.51 GB and 0.48 GB are reported for MHSA and SummaryMixing with 10-second long sentences compared to  1.9 GB and 0.8 GB at 120 seconds. Hence the delta increases from  6.25\% to 137\% demonstrating that SummaryMixing is much more efficient than MHSA with longer sentences enabling the use of infinite left-context.

\section{Conclusion}

This article extended SummaryMixing, a linear-time complexity alternative to self-attention, to streaming speech recognition with a streaming and non-streaming conformer transducer. The conducted experiments and analysis show that compared to self-attention, SummaryMixing leads to faster training, halves the required on-board memory for transcribing long utterances while keeping an infinite left context, and exhibits a constant real-time factor at decoding time both on CPU and GPU while reaching lower error rates on the selected streaming and non-streaming tasks.

\bibliographystyle{IEEEtran}
\bibliography{mybib}

\end{document}